\documentclass[english,aps,prl,twocolumn,oneside,floatfix]{revtex4}
\usepackage[T1]{fontenc}
\usepackage[latin9]{inputenc}
\usepackage{float}
\usepackage{graphicx}
\usepackage{amssymb}

\makeatletter

\usepackage{babel}
\makeatother

\begin{document}

\title{Direct Wigner tomography of a superconducting anharmonic oscillator}

\author{Yoni Shalibo$^{1}$, Roy Resh$^{1}$, Ofer Fogel$^{1}$, David Shwa$^{1}$,
Radoslaw Bialczak$^{2}$, John M. Martinis$^{2}$ and Nadav Katz$^{1}$}

\affiliation{$^{1}$Racah Institute of Physics, The Hebrew University of Jerusalem,
Jerusalem 91904, Israel}

\affiliation{$^{2}$Department of Physics, University of California, Santa Barbara,
California 93106, USA}

\begin{abstract}
The analysis of wave-packet dynamics may be greatly simplified when
viewed in phase-space. While harmonic oscillators are often used as
a convenient platform to study wave-packets, arbitrary state preparation
in these systems is more challenging. Here, we demonstrate a direct
measurement of the Wigner distribution of complex photon states in
an anharmonic oscillator - a superconducting phase circuit, biased
in the small anharmonicity regime. We test our method on both non-classical
states composed of two energy eigenstates and on the dynamics of a
phase-locked wavepacket. This method requires a simple calibration,
and is easily applicable in our system out to the fifth level.
\end{abstract}
\maketitle
The state of open quantum systems is frequently represented by a density
matrix \cite{Steffen2006,DiCarlo2010}. This description is useful
for a wide variety of systems, in which the probabilities and coherences
in some chosen basis are of interest. However, in systems with continuous
degrees of freedom, e.g. the relative position and momentum of atoms
in a molecule, wavepackets are often formed and therefore phase space
distributions are better suited to characterize the state and its
dynamics \cite{Schleich2001}. In particular, the Wigner distribution
offers direct information about expectation values and purity, and
provides a convenient framework to test the quantum-classical correspondence
\cite{Zurek2003,Katz2008}. In addition, since the Wigner distribution
holds complete information about the state, it can be directly transformed
into a density matrix and therefore its measurement is useful for
quantum state tomography \cite{Nielsen2004} as well. Numerous experiments
measured the Wigner distribution of harmonic systems using an auxiliary
qubit \cite{Bertet2002,Hofheinz2009}, or via heterodyne detection
\cite{Eichler2011}. While anharmonic systems exhibit a wider variety
of phenomena, a full quantum state reconstruction has so far been
limited to atomic and molecular systems \cite{Dunn1995,Weinacht1998,Skovsen2003,Hasegawa2008}.

In this letter, we report on a direct measurement of the Wigner distribution
in the Josephson phase-circuit, a superconducting anharmonic oscillator.
Our method utilizes simple tomography pulses, and as opposed to standard
state tomography, does not require individual calibration of the pulses.

Measuring the Wigner distribution in an anharmonic system poses several
challenges. First, the phase of each level $\phi_{n}$ in the rotating
frame advances in an increasing rate with $n$, causing wavepackets
to disperse in phase space during the tomography pulse. For example,
in a cubic potential this rate is given by $\dot{\phi}_{n}\thickapprox\beta n(n-1)/2$,
where $\beta=2\pi\left(f_{10}-f_{21}\right)$ is the anharmonicity
and $n=0,1,2,...$. Second, in our system it is impossible to measure
the probability distribution in the phase space quadratures directly,
and therefore the prominent method for phase-space tomography is measurement
of the parity after a coherent displacement \cite{Hofheinz2009}.
In order to achieve an approximate displacement operation, one has
to apply a pulse which is simultaneously resonant with all the transitions
within the measured subspace.  We find that both restrictions can
be practically met by both reducing the anharmonicity and applying
sufficiently short tomography pulses.

\begin{figure}
\centering{}\includegraphics[bb=30bp 20bp 265bp 215bp,clip,width=8cm]{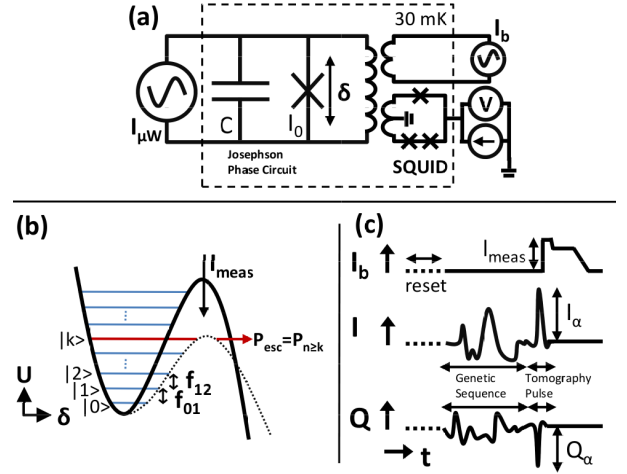}\caption{Experimental techniques. (a) The Josephson phase-circuit diagram.
(b) Measurement of the occupation probabilities using short pulses
in bias $I_{meas}$ that cause selective tunneling of states $n<k$
which are detected later with an on-chip SQUID. (c) Time sequence
of the Wigner tomography measurement of Fock-type superpositions.}
\end{figure}

Our system, the Josephson phase circuit (see Fig. 1a), is a superconducting
LC oscillator with a weak anharmonicity, originating from a Josephson
junction \cite{Clarke2008}. The potential energy of the system is
one-dimensional and has the form of a double-well as a function of
the phase difference $\delta$ across the junction. Using an external
bias current $I_{b}$ we tune the anharmonicity and measure the occupation
probabilities of the energy levels inside the smaller potential well
(see Fig. 1b). At large anharmonicity ($\beta/2\pi f_{01}\sim0.02$),
the lowest two levels are used for qubit operations via resonant microwave
pulses $I_{\mu w}$ that yield high-fidelity gates of only a few nanoseconds
in duration \cite{Lucero2010}. At small anharmonicity ($\beta/2\pi f_{01}\sim0.002$),
many more levels are accessible within a practical bandwidth and wave-packets
can be easily generated \cite{Shalibo2012}. In both regimes the state
is measured in the energy basis by applying a short pulse in $I_{b}$
that causes selective tunneling of the probability in occupied levels
$n>k$, where $k$ depends on the height of the potential barrier,
set by the pulse amplitude. Tunneling events are detected using an
on-chip superconducting quantum interference device (SQUID), and for
a set of measurement pulse amplitudes the occupation probabilities
are reconstructed from the tunneling probabilities \cite{Shalibo2012}.
While standard state tomography (SST) can be used to measure the full
density matrix of the lowest levels, in practice this method requires
complicated calibration procedures for states with more than two levels,
and is severely limited by decoherence. Our method requires only a
single calibration measurement for all the tomography pulses, and
can be scaled up to larger Hilbert spaces using optimized system parameters.

Ideally, the Wigner distribution is proportional to the expectation
value of the parity operator after a coherent displacement \cite{Royer1977}.
Due to the finite anharmonicity in our system, we use short, gaussian-shaped
resonant pulses as an approximate coherent displacement, while working
at a weak anhramonicity. We control the phase and size of the complex
displacement $\alpha$ by setting the phase and amplitude of the microwave
pulse, using the relation $\alpha=-\left(1/2\right)\int\Omega(t)dt$
for a harmonic system, where $\Omega(t)$ is the time-dependent, gaussian-shaped
Rabi amplitude. For our experimental parameters \cite{FN1}, the pulse
is simultaneously resonant with many transitions between consecutive
levels, $\sim6$ of which are subject to an amplitude variation of
less than 10\,\% from the peak amplitude \cite{Supp}. For a given
anharmonicity $\beta$, a resonant pulse can be well approximated
by a harmonic displacement in the limit $\beta T\left|\alpha\right|m^{2}/4\ll1$,
where $\left|\alpha\right|$ is the size of the applied displacement,
$m$ is the maximal occupied level after a displacement and $T$ is
the pulse duration \cite{Supp}. This condition limits the maximal
displacement to be well below the size of the distribution in phase
space ($\left|\alpha_{max}\right|\ll1$), for initial states occupying
up to 5 of the lowest levels. However, we find in simulation that
while the phases of the displaced state are very sensitive to the
above condition, the probabilities are not, and therefore our approximation
remains accurate for much larger displacements. We find that for our
experiment parameters, up to 5 levels can be measured accurately.
The expectation value of the parity operator is calculated from the
measured occupation probabilities and is given by, $\alpha=\left(2/\pi\right)\sum_{n}\left(-1\right)^{n}P_{n}$
\cite{Royer1977}. To test the effect of this pulse on our system,
we initiate our system in the ground state and measure the occupation
probabilities immediately after a short microwave pulse of total area
$\alpha$. Figure 2 shows the results of this measurement as a function
of state number $n$ and $\alpha$ (Fig. 2a), compared with the expected
probabilities in a harmonic oscillator (Fig. 2b), $P(\alpha,n)=\left(1/n!\right)\mathrm{exp}(-\left|\alpha\right|^{2})\left|\alpha\right|^{2n}$
. As expected, the probability distribution in $n$ is narrower compared
to the harmonic system for higher amplitudes due to our finite bandwidth.
To compare the data and theory quantitatively, we plot a histogram
of the distribution (Fig. 2c,d) for $\alpha=-1.3$ and $\alpha=2.2$.
At $\alpha=-1.3$ our data fits well with a harmonic displacement.
At the largest displacement values ($\alpha=2.2$) the deviation from
harmonicity becomes more apparent as expected \cite{Supp}.

\begin{figure}
\centering{}\includegraphics[bb=0bp 30bp 240bp 230bp,width=8cm]{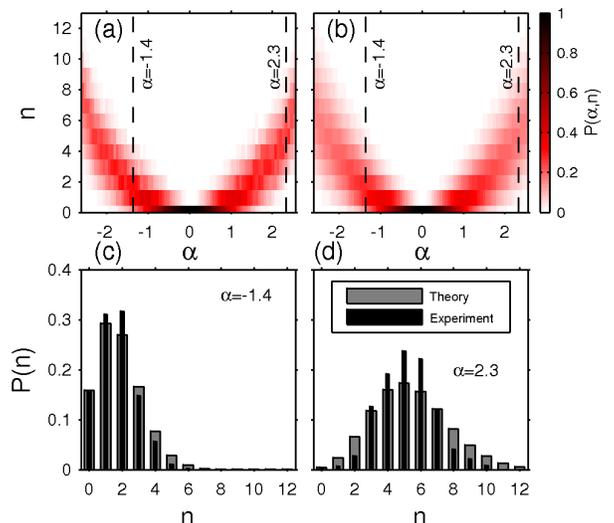}\caption{Tomographic pulse. Occupation probabilities for levels $n\leq12$,
as a function of $\left|\alpha\right|$ in the experiment (a) and
theory (b). (c) and (d) show a histogram along the dashed lines shown
in (a) and (b).}
\end{figure}

To benchmark our method, we apply it on a set of superpositions of
eigenstates of the system, $\left|\psi_{l}\right\rangle =\left(\left|0\right\rangle +e^{i\phi}\left|l\right\rangle \right)/\sqrt{2}$
\cite{FN3}. These states have a simple structure in phase space and
cannot disperse since they contain one measurable phase that results
in only a free rotation. As we operate our system in the small anharmonicity
regime, preparing such states with high fidelity requires longer pulse
sequences. Due to our short decay and coherence times ($T_{1}=120$\,ns,
$T_{2}>150$\,ns) we apply optimized control sequences to reduce
decoherence effects to a minimum (see Fig. 1c). We use a genetic optimization,
where we define the population overlap $\chi=\overrightarrow{P_{ideal}}\cdot\overrightarrow{P_{meas}}$
and optimize it with a feed-back loop \cite{Supp}. For a desired
state $\left|\psi_{l}\right\rangle $, we set $\overrightarrow{P_{ideal}}=\left(0.5,0,..,0.5,0,...\right)$
and use our measured occupation probabilities for $\overrightarrow{P_{meas}}$.
The results of the optimization algorithm are summarized in Table
1.

\begin{table}
\begin{centering}
\begin{tabular}{|c|c|c|c|c|c|c|c|c|}
\hline
$l$ & $P(\left|0\right\rangle )$  & $P(\left|1\right\rangle )$ & $P(\left|2\right\rangle )$ & $P(\left|3\right\rangle )$ & $P(\left|4\right\rangle )$ & $P(\left|5\right\rangle )$ & $\tau$ (ns) & $\chi$ (\%)\tabularnewline
\hline
\hline
1 & 0.52 & 0.47 & 0.01 & 0 & 0 & 0 & 15 & 99.8\tabularnewline
\hline
2 & 0.69 & 0.05 & 0.24 & 0.02 & 0 & 0 & 30 & 93.4\tabularnewline
\hline
3 & 0.58 & 0.03 & 0.11 & 0.27 & 0.01 & 0 & 25 & 90.5\tabularnewline
\hline
4 & 0.70  & 0.01  & 0.03  & 0.07  & 0.17  & 0.02 & 40 & 88\tabularnewline
\hline
\end{tabular}
\par\end{centering}

\caption{Results of the optimization algorithm.}
\end{table}

\begin{figure*}
\centering{}\includegraphics[bb=20bp 14bp 470bp 272bp,clip,width=17.2cm]{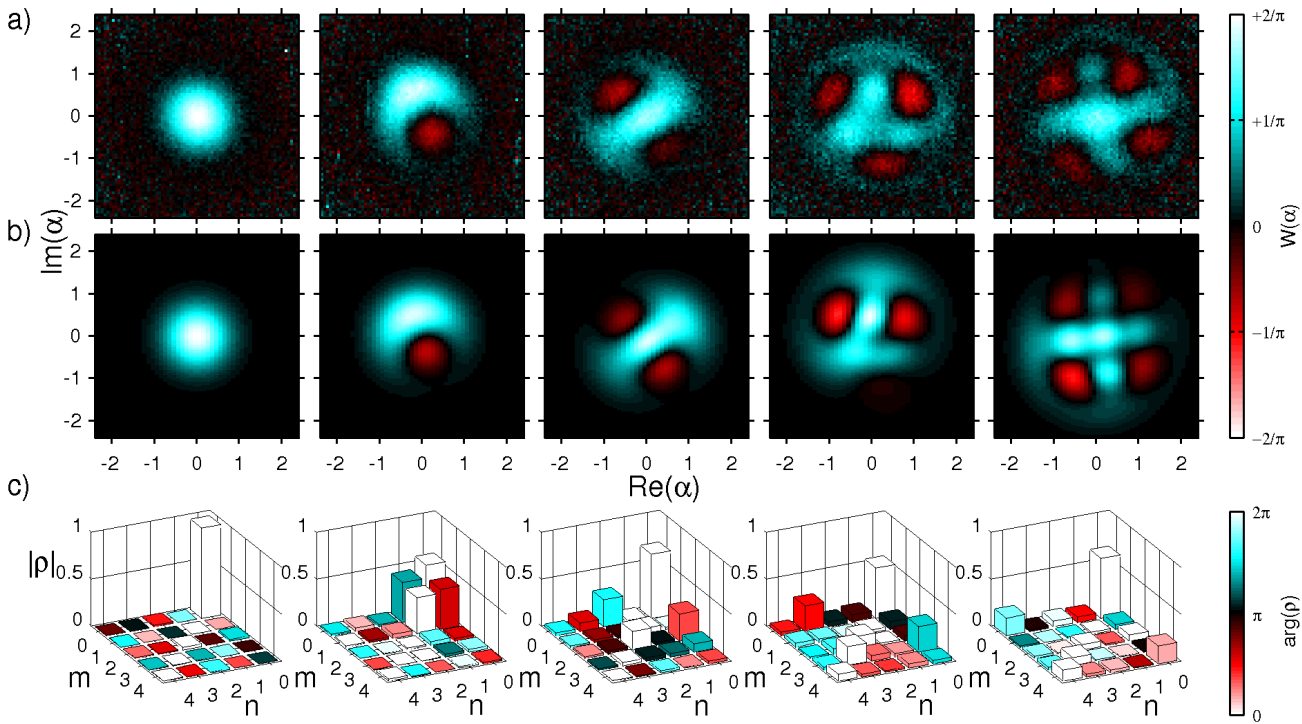}\caption{Superposition of Fock-type states. (a) Wigner tomography of genetically
optimized superposition of fock-type states and (b) calculated Wigner
distributions of these states, using level occupations from table
1 and phases of the extracted density matrices. (c) Extracted density
matrices from the measurements shown in (a).}
\end{figure*}

As expected, $\chi$ decreases for a larger $n$, mostly due to our
relatively short preparation pulse duration $\tau$. To increase the
value of $\chi$ for these states we increase $\tau$, at the expense
of the states' purity and optimization run-time. Our optimization
algorithm does not take into account phases, however for sufficiently
high values of $\chi$, the Wigner image is dominated by only one
phase, as can be seen in Fig. 3. In Fig 3a we plot the results of
the tomography measurement on the ground state, and states $\left|\psi_{l}\right\rangle $
described in Table 1. These are compared with the expected images
(Fig. 3b), calculated using the measured occupation probabilities,
obtained independently by the optimization algorithm \cite{FN4}.
In Fig. 3c we plot the density matrices of the measured states, extracted
using the harmonic oscillator wavefunctions. The diagonal elements
in the extracted density matrices agree well with the measured occupation
probabilities up to level $n=4$. However, simulation shows that off-diagonal
elements deviate increasingly for $n>2$ \cite{Supp}. We find that
the origin of these errors is primarily the significant decay and
decoherence that occur at highly excited levels during the tomography
pulse, and secondarily the finite bandwidth of our tomography pulse.
The errors in the measured Wigner distribution can be further reduced
with currently available higher coherence samples \cite{Supp}.

In general, for any finite set of repetitions, the amount of information
that is gained about the density matrix is smaller in a Wigner measurement
relative to SST (assuming equal number of repetitions and tomography
pulses). This is due to the large overlap between the unitary operations
associated with the Wigner tomography pulses, compared to the nearly
orthogonal operations in a typical SST measurement. However, we find
that the overhead in the number of repetitions in a Wigner measurement
relative to SST with similar uncertainties is of order unity \cite{Supp},
thereby making the Wigner measurement a practical alternative for
measuring the density matrix.

\begin{figure*}[t]
\begin{centering}
\includegraphics[bb=20bp 14bp 480bp 384bp,clip,width=17.2cm]{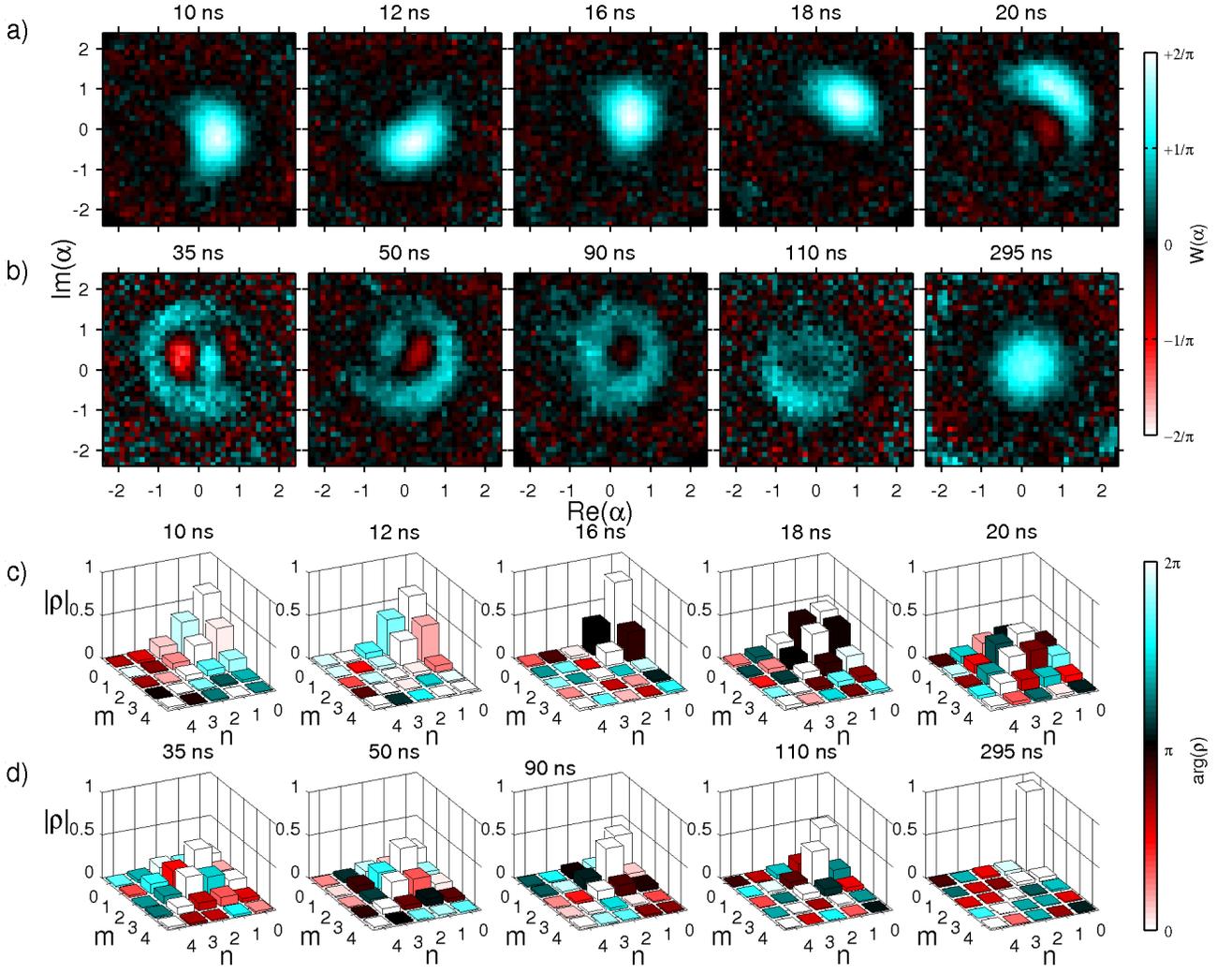}
\par\end{centering}

\centering{}\caption{Dynamics of a phase-locked wave-packet. (a) Wigner tomography during
a chirp ($T=20$\,ns, $\Omega/2\pi=66$\,MHz, $f_{in}-f_{01}=320$\,MHz,
$f_{fin}-f_{01}=-50$\,MHz), and (b) during drive-free evolution.
(c) and (d) show the extracted density matrices for (a) and (b) respectively.}
\end{figure*}

Non-dispersive wavepackets are particularly interesting to measure
in phase space, due to phase-locking between the system and the drive
that can take place within certain conditions \cite{Fajans2001}.
In a system with negative anharmonicity ($f_{01}>f_{12}$), a negative
frequency-chirped drive ($\dot{f}_{drive}<0$) can cause phase-locking
at sufficiently large amplitudes, depending on the chirp rate and
anharmonicity. To measure this effect, we initiate the system in the
ground state and apply a negative frequency chirp, with a drive amplitude
above the phase-locking threshold and a final frequency centered close
to the transition frequency $f_{23}$. The chirp's temporal length
and bandwidth $\left|f_{fin}-f_{in}\right|$ are chosen to be short
(20\,ns) and large (600\,MHz) respectively, in order have a broad
excitation of states and for the excitation to be adiabatic \cite{Shalibo2012}.
Figure 4 shows the result at selected times along the chirp (Fig.
4a), and after the drive is turned off (Fig. 4b). The axes in the
images are rotated at each time frame to fit the rotating frame of
the drive. During the chirp, we see a displacement of the ground state
distribution that gradually acquires a constant phase as the drive
crosses the linear resonance ($f=f_{01}$). This happens, as expected
around $t=16$\,ns and the shape of the wavepacket becomes crescent-like.
After the drive is turned off ($t=21$\,ns), phase-locking is lost
and the wavepacket disperses due to the finite anharmonicity. At $t=35$\,ns,
the state has completely dispersed, however it still contains significant
coherences, as indicated by the negative values in the Wigner plot
and the large off-diagonal elements in the extracted density matrix
(Fig. 4d). The state then dephases into a ring shape which shrinks
in radius towards the ground state. To track the decoherence dynamics
in this experiment, we extract the state's purity evolution directly
from the Wigner distribution: $\wp=\pi\int d\overrightarrow{\alpha}\left|W(\overrightarrow{\alpha})\right|^{2}$
\cite{Schleich2001}. The result is shown in Fig. 5 (red circles),
and compared with a simulation (solid line). As expected, the purity
remains high during the chirp, and then decays as a result of decoherence.
The purity reaches a minimum and then ascends towards unity with an
exponential rate, consistent with the measured decay time ($T_{1}=120$\,ns)
and in accordance with simulation.

\begin{figure}
\centering{}\includegraphics[bb=15bp 14bp 248bp 125bp,clip,width=8cm]{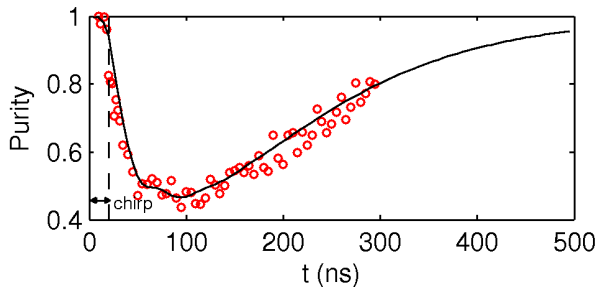}\caption{State purity. Extracted state purity $\wp$ from experiments (red
circles) and simulation (solid line) during chirp and decay. The simulation
includes the effect of the decay and dephasing and agrees with the
measured decay time of $T_{1}=120$\,ns.}
\end{figure}

In conclusion, we demonstrate a method of measuring the state of a
multi-level system with a small anharmonicity in phase space. Using
this method we are able to accurately extract the density matrix of
up to 5 levels, limited by systematic errors caused by decoherence
and finite bandwidth. This represents a significant improvement on
standard state tomography which is difficult to implement in this
subspace. The tunability of the phase circuit offers an approach of
improving the accuracy of such a measurement at \emph{larger} anharmonicity,
by quickly changing the bias $I_{b}$ after state preparation to the
small anharmonicity regime where the tomography pulse is approximately
a harmonic displacement. Finally, we point out that the overhead in
the number of measurements required to extract the density matrix
in a Wigner measurement is relatively small, and in addition one gains
global information about the state faster than in standard state tomography.

This work was supported by ISF grant 1248/10 and BSF grant 2008438.


\begin{thebibliography}{References}
\bibitem[18]{FN1} We use a 1.6\,ns FHWM pulse and an anharmonicity
$\beta/2\pi=20$\,MHz.
\bibitem[20]{FN3}$\phi$ is an unknown phase that can be determined
via Wigner tomography. Since our optimization algorithm is only sensitive
to state populations and not to phases, we cannot know it in advance.
\bibitem[21]{FN4}Due to the non-negligible occupation probabilities
we get at larger $l$ states, we add phases to the Wigner distribution
to account for additional features in the image
\end{thebibliography}
\end{document}


\title{Direct Wigner tomography of a superconducting anharmonic oscillator}

\author{Yoni Shalibo, Roy Resh, Ofer Fogel, David Shwa, Radoslaw Bialczak,
John M. Martinis and Nadav Katz}

\maketitle
\textbf{\large Supplementary Information}{\large \par}

\textbf{Materials and methods}. The Josephson phase circuit \cite{Martinis2002}
used in the experiment has the following design parameters: critical
current $I_{0}\approx1.5\,\mu$A, capacitance C$\approx$1.3\,pF
and inductance L$\approx$940\,pH. The qubit has a tunable frequency
$f_{01}$ in the 6-9\,GHz range. During the experiment the device
is thermally anchored to the mixing chamber of a dilution refrigerator
at 30\,mK, where thermal excitations of the qubit are negligible.

We use a custom built arbitrary waveform generator (AWG) having a
fast (1\,ns time resolution), 14-bit digital-to-analog converter
to shape the microwave pulses which control the quantum state of the
circuit. We control the phase and amplitude of the drive by modulating
a high-frequency oscillator of frequency $f_{\mathrm{LO}}$ with the
AWG, using an IQ-mixer. The modulation signals are fed into the I
and Q ports of the IQ-mixer to give an amplitude $\sqrt{I(t)^{2}+Q(t)^{2}}$
and a relative phase $\phi(t)=\arctan\left(Q(t)/I(t)\right)$ at its
output.

\textbf{Harmonic response.} As indicated in the manuscript, the bandwidth
of the tomography pulse must be large relative to the anharmonicity,
for a nearly harmonic response. This is easily understood by looking
at the position of the transitions in the spectral domain, relative
to the local oscillator frequency (see Fig. \ref{FiniteBW}). As seen
in the figure, the effective drive amplitude of the lowest 6 transitions
is the same, to within a 10\,\% variation. This is consistent with
our data and simulations (see Sec. II), showing small systematic errors
within the corresponding subspace. In principle, more transitions
can fit within this region by positioning the local oscillator lower
in frequency.

\renewcommand{\thefigure}{S\arabic{figure}}

\begin{figure}
\centering{}\includegraphics[bb=0bp 10bp 285bp 230bp,clip,width=0.7\columnwidth]{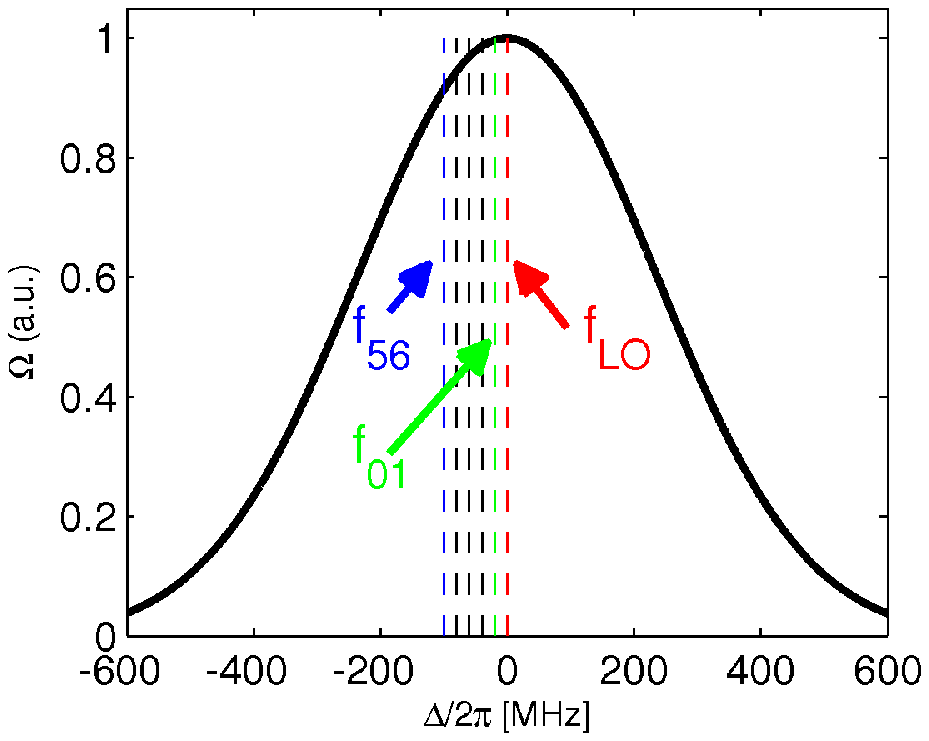}\caption{\label{FiniteBW} Tomography pulse envelope in the frequency domain.
Dashed lines are the lowest five transitions $f_{n,n+1}$ and the
local oscillator frequency position in the experiment. The pulse envelope
(solid line) is the normalized fourier transform of a 1.6\,ns FWHM
gaussian function.}
\end{figure}

\renewcommand{\thefigure}{S\arabic{figure}}

\begin{figure}
\centering{}\includegraphics[bb=145bp 135bp 550bp 405bp,clip,width=0.6\columnwidth]{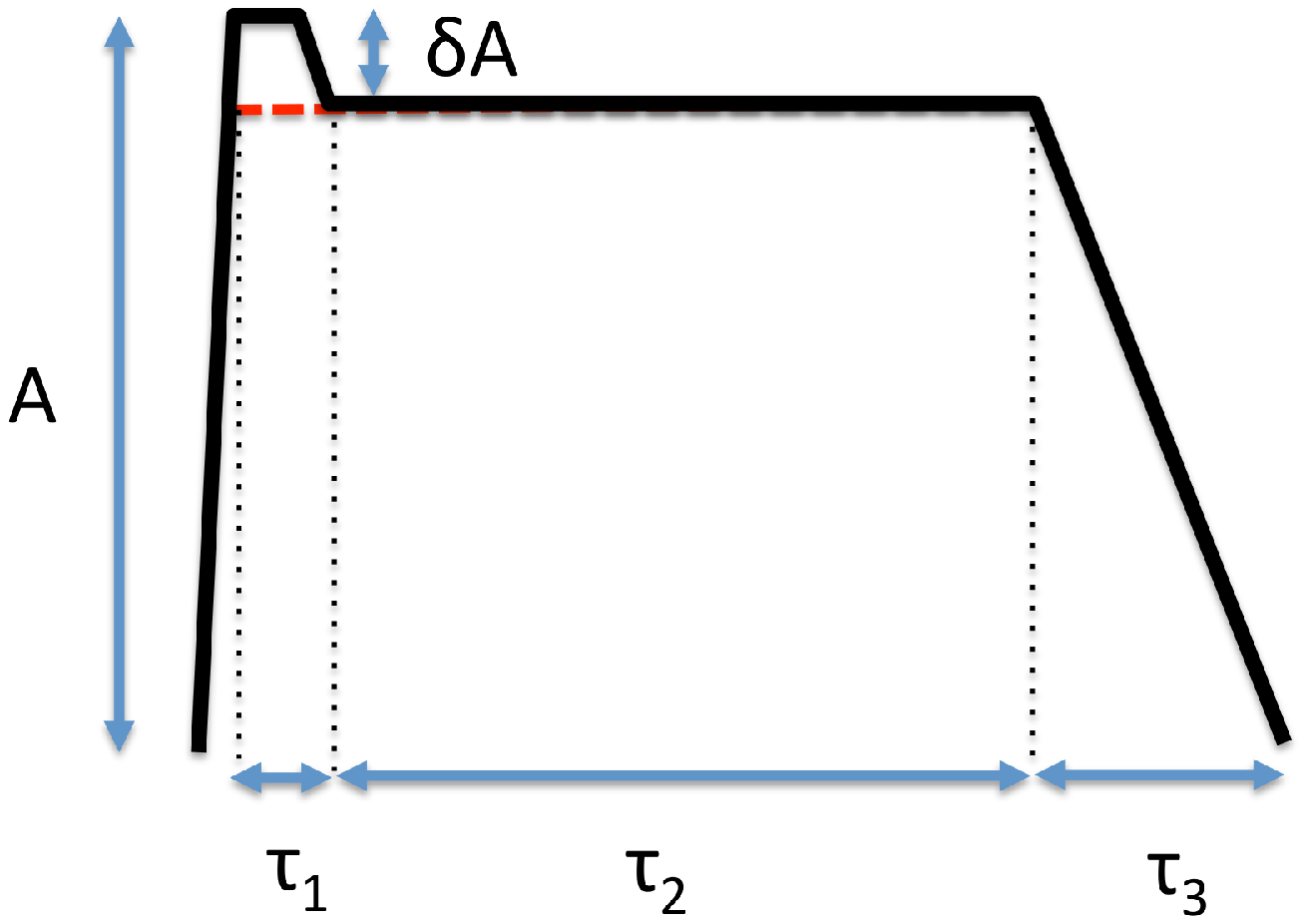}\caption{\label{mp} Measurement pulse used in the experiment. All the pulse
parameters except for the total amplitude $A$ are fixed.}
\end{figure}

\textbf{Populations measurement. }The measured escape probabilities
due to a set of experiments with different measurement pulses are
converted into level occupation probabilities using the measured escape
curves of the lowest levels in the well \cite{Shalibo2012}. In this
experiment, we used a novel measurement pulse shape, to reduce distortions
of the escape curves of higher levels. This measurement pulse is shown
in Fig. \ref{mp}. Typically, one must use a pulse with a slowly decaying
end, in order to reduce the effect of population re-trapping in the
original well after the pulse ends. The effect of re-trapping becomes
more prominent as the anharmonicity is reduced due to the corresponding
deepening of the potential well, and therefore requires increasing
both $\tau_{2}$ and $\tau_{3}$ to eliminate re-trapping. At the
working bias point used in our experiment, we use $\tau_{2}=25$\,ns
and $\tau_{3}=15$\,ns. These parameters are non-negligible relative
to the decay time and cause artificial increase of the extracted population
at lower levels, which in turn distorts the Wigner image. To avoid
this effect we add a small step in amplitude at the beginning of the
pulse ($\tau_{1}=1$\,ns) to preselect the population we want to
escape, and then reduce the amplitude by an amount $\delta A$. Even
the small difference $\delta A$ is sufficient to protect undesired
population at lower levels from tunneling out during the long waiting
time in which the escaped population decays in the external well.

\textbf{Density matrix fit. }In the Wigner tomography experiments
we extract the density matrix from the populations of the displaced
states \cite{Hofheinz2009}. We use 200 homogeneously distributed
random displacements within a $\left|\alpha\right|<2$ circle to fit
the density matrix, while restricting the density matrix to a $6\times6$
subspace. It should be noted that both the measured Wigner distribution
and the extracted density matrix represent the state after a rotation
that occurs during the tomography pulse. To get more accurate phases,
one can apply an inverse propagator on the density matrix $U=\exp(-iH_{0}\Delta t/\hbar)$,
where $H_{0}$ is the drive-free Hamiltonian and $\Delta t$ is the
effective pulse length for the rotation.

\section{State preparation by genetic optimization}

Finite anharmonicity makes it possible to prepare states, composed
of arbitrary superpositions of eigenstates within our system. However,
fourier broadening of the drive, together with power broadening of
the transition energies causes nontrivial excitation at relatively
small anharmonicities. Adding the short decay and coherence time,
it becomes challenging to prepare a desired, yet pure state. We solve
this difficulty, by optimizing the state produced with a feedback
from the experiment. Our target state in the optimization algorithm
is a superposition of eigenstates of our system: $\left|\psi_{l}\right\rangle =\left(\left|0\right\rangle +e^{i\phi}\left|l\right\rangle \right)/\sqrt{2}$.
Using the measured probabilities we are able to optimize such a superposition
up to an unknown phase $\phi$ that causes only a free rotation of
the measured Wigner function.

Our optimization algorithm is based on guided evolution. It is constructed
in the following steps:

(a) We define a set of $N_{G}$ pulse sequences (genomes). Each genome
is a sequence of $N_{t}$ complex numbers that represent the amplitude
and phase of the drive's pulse envelope at each time step, with 1\,ns
time resolution. In addition, each genome is associated with a population
overlap, defined as: $\chi=\overrightarrow{P_{ideal}}\cdot\overrightarrow{P_{meas}}$,
where $\overrightarrow{P_{ideal}}$ and $\overrightarrow{P_{meas}}$
are the desired and measured population vectors . The genomes are
initialized with random complex numbers, each having a maximal amplitude
$\Omega_{max}$, and $\chi=0$.

(b) At each iteration of the algorithm (generation), we assume that
the current set of genomes is sorted in order of decreasing $\chi$.
The upper $N_{a}$ genomes are kept, and the next $N_{a}$ genomes
are replaced by a combination of the upper $N_{a}+1$ genomes: $\overrightarrow{G_{N_{a}+k}}=\overrightarrow{P}(G_{k},G_{k+1})+\overrightarrow{\epsilon}$,
where the function $\overrightarrow{P}(x,y)$ randomly selects an
amplitude between the sets $x$ and $y$ at each time step and $\overrightarrow{\epsilon}$
is a set of random complex numbers with a small amplitude $\epsilon_{max}$
(noise). The bottom $N_{G}-2N_{a}$ genomes are replaced by random
complex numbers with an amplitude $\Omega_{max}$.

(c) The bottom $N_{G}-N_{a}$ genomes of the next generation are then
applied and the resulting probabilities $\overrightarrow{P_{meas}^{k}}$
for each genome are measured. After calculating $\chi$ of each, we
sort our new set $G$ in order of decreasing $\chi$, and mark genome
sequences that have (potentially) higher $\chi$ than the current
optimal $\chi$. For each genome having a potentially higher $\chi$
we repeat the measurement $N_{rep}$ times, recalculate $\chi$ and
reposition them in the set $G$. Step (b) is run again.

\textbf{Fidelity measure.} Our population overlap $\chi$ measures
the distance between two states. There are many possibilities of defining
the fidelity, and in general each gives a different bias for the algorithm
towards finding a specific family of states. In our definition for
example, $\chi$ is highly sensitive to the overall population that
exists outside the subspace that is populated in the target state.
However it is much less sensitive to the distribution among levels
that are populated in the target state. Our definition is particularly
useful for optimizing our target states $\left|\psi_{l}\right\rangle $
for a Wigner measurement, because we wish to eliminate the population
of states other than $\left|0\right\rangle $ and $\left|l\right\rangle $.

There are two important factors that limit the performance of any
optimization algorithm in experimental systems: shot noise, and drifts
in the physical parameters of the system.

\textbf{Shot noise.} Any finite number of repetitions results in some
uncertainty in the measured probabilities. To determine $n$ occupation
probabilities in the final state we measure the tunneling probabilities
after $n$ measurement pulses having different amplitudes. In a typical
optimization algorithm, we repeat the measurement of a single tunneling
probability 900 times. Statistical analysis gives a typical uncertainty
(standard deviation) of $\sim$2\,\% in $\chi$ for sufficiently
high values of $\chi$ ($\chi\gtrsim80$\,\%) and therefore one cannot
distinguish between two genomes with a $\chi$ difference which is
smaller than $\sim$3\,\%. We therefore expect the algorithm to run
significantly more slowly at $\chi$'s approaching unity, due to false
increase of $\chi$. To increase the efficiency of the algorithm without
increasing the number of repetitions of each measurement, we repeat
the measurement only for genomes with potentially increased $\chi$,
as described in step (c).

\textbf{Drifts.} The response of the system to the application of
a given genome can change in time due to drifts in its physical parameters.
We correct for drifts in the energies by performing a spectroscopic
measurement of the qubit transition frequency every 10 minutes and
adjusting the flux bias to restore the qubit frequency to its original
value. In addition, we correct for drifts in the offset voltages applied
to the IQ mixer that eliminate leakage of the microwave drive while
it is turned off.

\renewcommand{\thefigure}{S\arabic{figure}}

\begin{figure}
\centering{}\includegraphics[bb=0bp 0bp 285bp 230bp,clip,width=0.8\columnwidth]{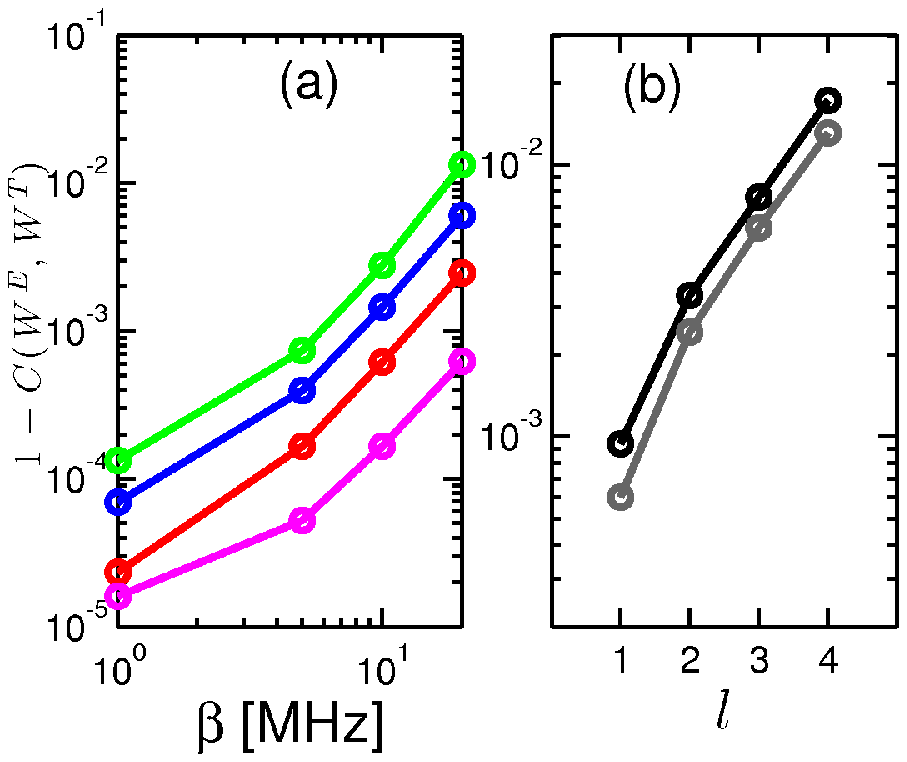}\caption{\label{WigSysErrs} Systematic errors in the Wigner distribution,
for initial states $\left|\psi_{l}\right\rangle =\left(\left|0\right\rangle +\left|l\right\rangle \right)/\sqrt{2}$,
where $l=1,2,3,4$. The cross-correlation deviation as a function
of the anharmonicity (a) and initial state. In (a), decoherence is
not included, and the magenta, red, blue, green lines correspond to
the initial state $l=1,2,3,4$ respectively. In (b) decoherence is
included and the anharmonicity is fixed to $\beta=20$\,MHz. The
black and grey lines correspond to decoherence parameters $T_{1}=150$\,ns,
$T_{2}=200$\,ns and $T_{1}=600$\,ns, $T_{2}=600$\,ns respectively.}
\end{figure}

\section{Systematic errors}

To quantify the errors in the measured Wigner function and density
matrix caused by finite anharmonicity and decoherence, we perform
numerical simulations . In our simulations, we propagate an anharmonic
system, initialized with a pure state $\rho=\left|\psi\right\rangle \left\langle \psi\right|$,
with resonant pulses. Each pulse is assumed to be resonant with the
first transition and has a gaussian envelope with the same width as
in the experiment. We use the rotating wave approximation for constructing
the Hamiltonian, and neglect corrections to the drive coupling beyond
the harmonic approximation, namely $\left\langle n\right|\delta\left|m\right\rangle =0$
for $n\neq m\pm1$ and $\left\langle n\right|\delta\left|m\right\rangle =\sqrt{n},\sqrt{n+1}$
for $m=n\pm1$. This results in the following time dependent unitary
propagator: $\mathrm{U}(t)=\exp\left(i\frac{\delta t}{2}\left[\left(\Omega(t)a^{\dagger}+\Omega(t)^{*}a\right)+\beta a^{\dagger}a(a^{\dagger}a-1)\right]\right)$,
where $\delta t$ is the time step in the simulation, $\Omega(t)$
is the time dependent drive envelope amplitude and $\beta=2\pi\left(f_{01}-f_{12}\right)$
is the anharmonicity. Decoherence and energy relaxation are taken
into account using quantum operations, assuming only two phenomenological
parameters: the energy relaxation time of the qubit $T_{1}$ and its
pure dephasing time $T_{2}$.

We expect to have negligible systematic errors due to finite anharmonicity
when the second term in the exponent of the propagator becomes negligible.
By separating the terms in the exponent to first order, and assuming
a constant drive amplitude $\Omega$ of total duration $T$, we can
approximate the propagator to $U\approx D(\alpha)\exp(-i\frac{T\beta}{4}[\left(\alpha a^{\dagger}-\alpha^{*}a\right),a^{\dagger}a(a^{\dagger}a-1)])$
where $D(\alpha)$ is the displacement operator and $\alpha\equiv i\Omega T$.
The second term can be considered negligible in the limit, $\left|\alpha\right|T\beta m^{2}/4\ll1$,
where $m$ is the maximal occupied state. For our experimental parameters
($\beta/2\pi=20$\,MHz, $T=1.6$\,ns), the second term can be neglected
only for $\left|\alpha\right|\ll1$, however, the error in the Wigner
distribution, obtained from the state populations after the pulse
are negligible even for $\alpha\approx2$, as we show in our analysis.

We plot the errors in the Wigner distribution and the extracted density
matrices, as a function of the anharmonicity $\beta$, initial state
and decoherence parameters. The density matrix is extracted, as in
the experimental data, using only the populations of the lowest 6
levels.

\renewcommand{\thefigure}{S\arabic{figure}}

\begin{figure}
\centering{}\includegraphics[bb=14bp 14bp 295bp 352bp,clip,width=0.8\columnwidth]{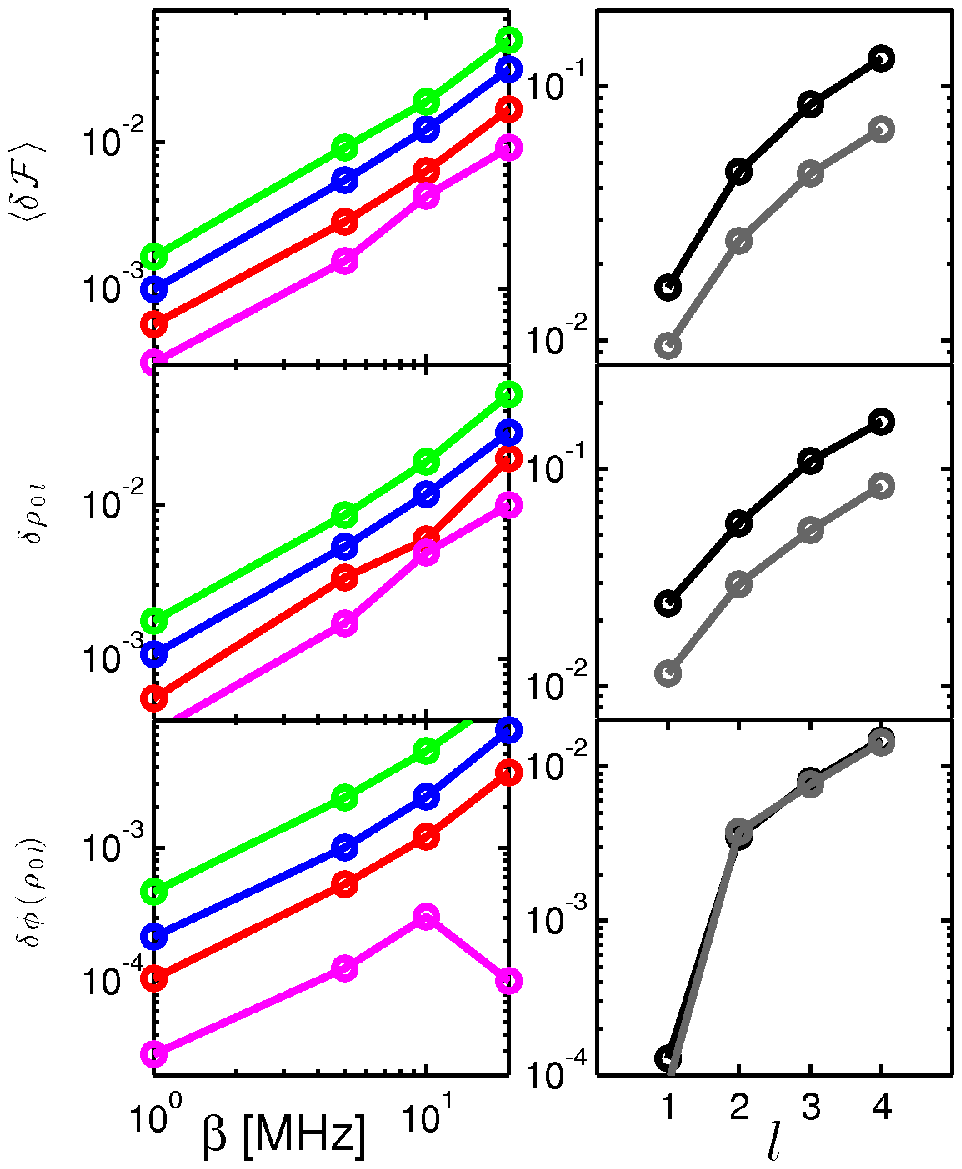}\caption{\label{WigDMSysErrs} Systematic errors in the extracted density matrix,
for initial states $\left|\psi_{l}\right\rangle =\left(\left|0\right\rangle +\left|l\right\rangle \right)/\sqrt{2}$,
where $l=1,2,3,4$. The left column of plots shows the fidelity error,
and errors in the off-diagonal matrix element $\rho_{0l}$ as a function
of anharmonicity, in the absence of decoherence. The magenta, red,
blue and green lines correspond to the initial state $l=1,2,3,4$
respectively. The same error measures are plotted in the right column,
as a function of initial state $l$, with decoherence included, at
$\beta=20$\,MHz. The black and grey lines correspond to decoherence
parameters $T_{1}=150$\,ns, $T_{2}=200$\,ns and $T_{1}=600$\,ns,
$T_{2}=600$\,ns respectively.}
\end{figure}

To quantify the error in the Wigner distribution, we calculate the
cross-correlation (at zero offsets) between the ideal Wigner distribution
of the initial state, and the one obtained from the expectation value
of the parity operator after a set of displacements. The cross correlation
is defined as $C(f(x,y),g(x,y))=\underset{x',y'}{\sum}\frac{\left(f(x',y')-\left\langle f\right\rangle \right)\left(g(x',y')-\left\langle g\right\rangle \right)}{\sqrt{\underset{x'',y''}{\sum}\left(f(x'',y'')-\left\langle f\right\rangle \right)^{2}\underset{x,y}{\sum}\left(g(x'',y'')-\left\langle g\right\rangle \right)^{2}}}$,
where $\left\langle f\right\rangle ,\left\langle g\right\rangle $
are the average values of $f,g$. The results are plotted in Fig.
\ref{WigSysErrs}. As seen in the figure, the error increases sharply
with both anharmonicity and maximal populated level $l$.

To quantify the error in the extracted density matrices, we use the
standard fidelity definition: $\left\langle \mathcal{\delta F}\right\rangle =1-Tr\sqrt{\sqrt{\rho^{F}}\rho^{I}\sqrt{\rho^{F}}}$,
where $\rho^{I}$ is the initial density matrix and $\rho^{F}$ is
the density matrix obtained from a fit to the populations of the displaced
states. We choose the initial density matrix $\rho^{I}$ to be $\left|\psi_{l}\right\rangle \left\langle \psi_{l}\right|$,
where $\left|\psi_{l}\right\rangle =\left(\left|0\right\rangle +\left|l\right\rangle \right)/\sqrt{2}$.
In addition, we calculate the error in the non-zero off-diagonal elements
of the density matrix using the following definitions: $\delta\rho_{0l}=\left|\left|\rho_{0l}^{F}\right|-\left|\rho_{0l}^{I}\right|\right|$
is the error in the amplitude of the matrix element, and $\delta\phi(\rho_{0l})=\left|\phi\left(\rho_{0l}^{F}\right)-\phi\left(\rho_{0l}^{I}\right)\right|/2\pi$
is the normalized error in the phase of the matrix element.

As seen in the Fig. \ref{WigDMSysErrs}, all the error measures are
negligible (smaller than $\sim0.05$) when decoherence is neglected.
However, when included, both the fidelity and the amplitude of the
off-diagonal elements have non-negligible errors. For currently available
samples having $T_{1}>600$\,ns, and correspondingly longer $T_{2}$,
the errors due to decoherence can be substantially reduced; In this
case, the errors become smaller than 0.1 in all the measures. As expected,
the phase error is quite insensitive to decoherence.

\section{Shot noise in Wigner tomography vs. standard state tomography}

\renewcommand{\thefigure}{S\arabic{figure}}

\begin{figure}
\centering{}\includegraphics[bb=14bp 17bp 302bp 178bp,clip,width=0.8\columnwidth]{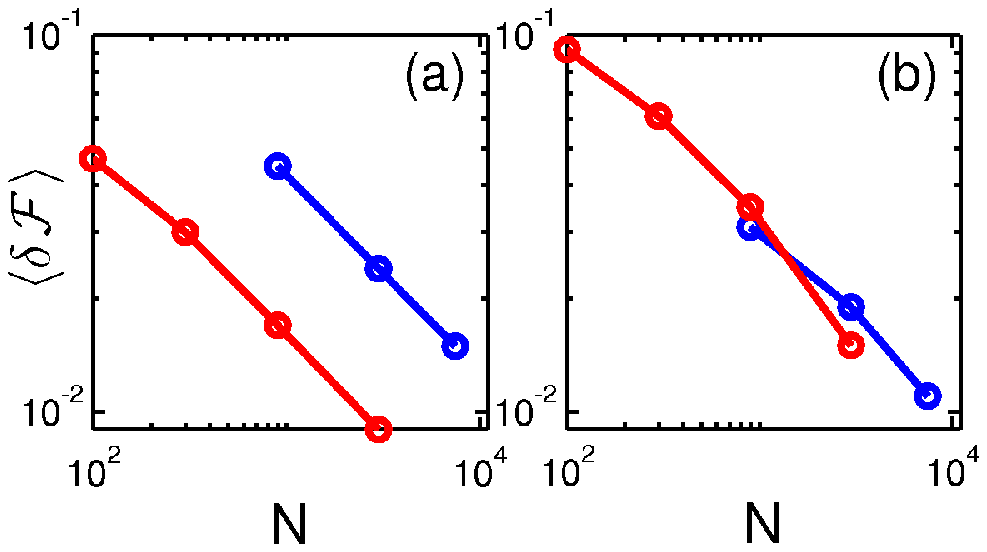}\caption{\label{WigVsSST} Error in the fidelity due to shot noise in standard
state tomography and WT. $\left\langle \mathcal{\delta F}\right\rangle $
vs. $N$ in WT (blue) and SST (red) of the initial states $\left|\psi\right\rangle =\left(\left|0\right\rangle +\left|4\right\rangle \right)/\sqrt{2}$
(a) and $\left|\psi\right\rangle =\left(1/\sqrt{5}\right)\protect\overset{4}{\protect\underset{k=0}{\sum}}\left|k\right\rangle $
(b). Both follow the $1/\sqrt{N}$ trend.}
\end{figure}

As pointed out in the manuscript, phase space is a convenient basis
to experimentally acquire global information about the state (e.g.
phase distribution, average energy etc.) \emph{fast}, on the expense
of accurate knowledge of the state in the eigenstate basis. Therefore,
when extracting the density matrix in the latter basis from a Wigner
measurement, one requires an excess number of measurements compared
to standard state tomography (SST) to achieve similar uncertainties.
It turns out that this is only true for a certain class of states.
For states that are dispersed in phase space, such as states composed
of a coherent superposition of a small number of eigenstates, SST
requires significantly less tomography pulses than Wigner tomography
(WT) to achieve a comparable error in the density matrix but the \emph{same}
number of pulses for states that are localized in phase space. The
overhead in the number of measurements is an important parameter from
an experimental standpoint, and therefore we perform numerical simulations
to calculate it. In the following we describe our simulation methods
for both cases.

\textbf{WT.} We start with a pure initial state $\left|\psi\right\rangle $.
We use the same method described in Sec. II to calculate the density
matrix after a coherent displacement. We keep the diagonal elements
of the final density matrix, for a set of $N_{W}$ random displacements
$\alpha$ that are uniformly distributed in the complex plane. For
each experiment (a particular displacement) we calculate an ensemble
of $M$ possible outcomes for the measurements of the diagonal elements,
assuming $r$ repetitions in the experiment, and a binomial distribution
for the escape probabilities, from which the diagonal elements $P(n)$
are calculated. For each outcome, we extract the density matrix and
calculate its fidelity $\mathcal{F}=Tr\sqrt{\sqrt{\rho^{tom}}\rho^{ideal}\sqrt{\rho^{tom}}}$.
We then calculate the average fidelity $\left\langle \mathcal{F}\right\rangle $
of the ensemble to find the experimental error $\left\langle \mathcal{\delta F}\right\rangle =1-\mathcal{\left\langle F\right\rangle }$
due to shot noise.

\textbf{SST.} We start with a pure initial state $\left|\psi\right\rangle $.
We construct a set of $N_{SST}=d^{2}$ orthogonal, unitary operations
$U_{j}$ to span a $d$-level subspace. The set of operations is chosen
to be the generators of $SU(n)$ for convenience. From the diagonal
elements of the rotated density matrices, we are able to extract the
expectation values of each generator $\left\langle U_{j}\right\rangle =Tr(\rho U_{j})$,
and therefore reconstruct the original density matrix: $\rho=\underset{j}{\sum}\left\langle U_{j}\right\rangle \mathbf{U}_{j}$
\cite{Thew2002}. As before, for each diagonal we calculate an ensemble
of $M$ possible measurement outcomes due to shot noise. From each
of the measurements in the ensemble, we extract the expectation values
of the operators $U_{j}$ and calculate the corresponding density
matrix. The average fidelity error of the ensemble of density matrices
is then evaluated.

Figure \ref{WigVsSST} shows the results of both simulations. We plot
$\left\langle \mathcal{\delta F}\right\rangle $ as a function of
$N=R$, where $R$ is the number of repetitions of the experiment
per tomography pulse. In the WT simulation we define $N$ relative
to the SST case: $N=R\left(\frac{N_{W}}{N_{SST}}\right)$, where $N_{W}$
is the number of tomography pulses in the Wigner simulation, and $N_{SST}$
is the number of tomography pulses in SST. While $N_{SST}$ is fixed,
we vary $N$ in the WT simulation by using many displacement pulses
while fixing $R$, and in the SST simulation we vary $N$ by changing
the statistics $R$. In all Wigner simulations we fix $R$ to be 900
(as in our experiment) and in the SST simulations we vary $R$ from
100 to 3000.

We see that for an initial state, composed of a coherent superposition
of only 2 states (Fig. \ref{WigVsSST}a), SST outperforms WT by a
factor of 8, in terms of the fidelity $\mathcal{F}$. In contrast,
for states composed of a coherent superposition of 5 eigenstates,
the amount of information that is extracted per pulse in SST and WT
is similar. This is because our chosen state is partially localized
in phase space, and therefore requires less displacement pulses to
extract its properties.